# FU-net: Multi-class Image Segmentation using Feedback Weighted U-net

*Mina Jafari[1(✉)], Ruizhe Li[1], Yue Xing[2], Dorothee Auer[2], Susan Francis[3], Jonathan Garibaldi[1], Xin Chen[1]*

[1]School of Computer Science, University of Nottingham, UK
Mina.jafari@nottingham.ac.uk
[2]School of Medicine, University of Nottingham, UK
[3]Sir Peter Mansfield Imaging Centre, University of Nottingham, Nottingham, UK

**Abstract.** In this paper, we present a generic deep convolutional neural network (DCNN) for multi-class image segmentation. It is based on a well-established supervised end-to-end DCNN model, known as U-net. U-net is firstly modified by adding widely used batch normalization and residual block (named as BRU-net) to improve the efficiency of model training. Based on BRU-net, we further introduce a dynamically weighted cross-entropy loss function. The weighting scheme is calculated based on the pixel-wise prediction accuracy during the training process. Assigning higher weights to pixels with lower segmentation accuracies enables the network to learn more from poorly predicted image regions. Our method is named as feedback weighted U-net (FU-net). We have evaluated our method based on T1-weighted brain MRI for the segmentation of midbrain and substantia nigra, where the number of pixels in each class is extremely unbalanced to each other. Based on the dice coefficient measurement, our proposed FU-net has outperformed BRU-net and U-net with statistical significance, especially when only a small number of training examples are available. The code is publicly available in GitHub[1].

## 1     Introduction

Image segmentation is a fundamental and crucial step in many image analysis tasks. In this paper, we focus on medical applications. From classical image segmentation methods (e.g. region growing) to more robust methods (e.g. level-set [1] and graph-cut [2]), various techniques have been proposed to achieve automatic image segmentation in a wide range of clinical problems. More recently, machine learning based methods have achieved superior performance against other traditional methods. It typically requires a training process, where a human-designed feature descriptor (e.g. SIFT[3] etc.) is applied to represent local image characteristics. Subsequently, the extracted features are used to train a classification model for pixel-level classification to achieve image segmentation.

---

[1] GitHub link: https://github.com/MinaJf/FU-net

Since 2012, based on the idea of convolutional neural network (CNN) proposed by LeCun et al. [4] and followed by a technological breakthrough that allows deeper neural networks to be trained [5], deep CNNs have demonstrated remarkable capabilities in performing classification, segmentation, object detection, and other image processing tasks [6, 7]. Briefly, the CNN-based methods recognize objects based on a multi-scale feature representation obtained by applying many convolutional filters and non-linear activation functions at different image scales. The parameters of the convolutional filters are automatically learned during the training process through iterative back propagation of the errors between the predicted outputs and the ground truth images. This enables an automatic feature learning and representation, which is the key advantage against classical machine learning methods that are based on manually designed features.

Many deep CNN based methods have been proposed to address image segmentation tasks. In earlier approaches, image segmentation is treated as a pixel-wise classification problem [8]. Deep CNN classification models are trained in a patch-based manner. These methods require millions of image patches for training and suffer from low computational efficiency in both training and testing stages. One of the latest state-of-the art methods (known as U-net [9]) is based on an end-to-end deep CNN architecture. It is trained more efficiently and requires fewer training samples than the patch-based models. Following on this pioneer work, several improvements and modifications have been proposed. For instance, Drozdzal et al. [10] added short skip connections in addition to the long skip connections in the U-net to improve training efficiency and segmentation accuracy. RU-net and R2U-net, proposed by Alom et al. [11], are based on U-net plus recurrent neural network and U-net plus the combination of recurrent neural network and residual network respectively. A nested U-net architecture called U-net++ is introduced in [12] that is proposed to replace the direct skip connections from encoder to decoder part by dense skip connections. A chain of multiple U-nets are utilized in LadderNet [13] to improve the flow of information.

For most multi-class image segmentation problems, the number of pixels in each class is different from each other which potentially leads to less accurate predictions for some classes than others. Additionally, some of the image regions are easier to be classified (i.e. higher segmentation accuracy) than others due to more distinct local image characteristics. It would be more efficient if the network can be dynamically adapted to learn from pixel locations with lower predicted accuracies during the training process. There are a few methods have been proposed to address these issues. Focal loss [14] is proposed to modify the cross entropy loss function for addressing the class imbalance problem. Similarly, online hard example mining method proposed by Shrivastava et al. [15] balances class samples by mining hard examples based on the loss values. Both methods focus on the problem of object classification, while the application to image segmentation has not been thoroughly investigated.

As the main contribution of this paper, we improve the U-net method by introducing a dynamically weighted cross-entropy loss function. The weight for each pixel is calculated based on the predicted accuracy in each iteration. The pixel locations with higher prediction accuracies are assigned with lower weights, and vice versa. This enables the network to learn more from poorly predicted image regions. We name our proposed method as feedback weighted U-net (FU-net). We demonstrate the effectiveness of the FU-net using a challenging brain magnetic resonance image (MRI) dataset with extremely unbalanced classes as well as different numbers of training samples.

## 2      Methodology

### 2.1      Network Architecture

The U-net proposed by Ronnebergeret et al. [9] is based on convolutional neural network, and consists of a contracting path and an expansive path. In the contracting path, each layer consists of two 3×3 convolutions (Conv), and each convolution is followed by a rectified linear unit (ReLU) as illustrated in Fig. 1(a). The feature map in the next successive layer is a down-sampled version of the output from the previous layer by using a max pooling of stride 2. Due to the down-sampling process, only very abstracted information remained at the end of the contracting path. To capture and rebuild the spatial context, a decoding path is required. In the expansive (decoding) path, the output feature map in each layer is up-sampled using 2×2 up-convolution with halved number of feature channels in the previous layer. Each layer also has two 3×3 convolutions, and each followed by a ReLU. Additionally, there are some concatenation operations to combine feature maps from the contracting layers to the corresponding expansive layers. 1×1 convolution is used in the final layer to convert the dimension of feature maps to the number of classes. Subsequently, softmax function [16] is applied to map the output value of each pixel to the range of [0, 1]. In the U-net paper [9], the authors proposed a weighted cross entropy loss function $E$ for parameter optimization that is expressed in equation (1).

$$E = \sum_{x \in \Omega} w(x) \, log(p_{l(x)}(x)) \qquad (1)$$

where $p_{l(x)}(x)$ is the predicated probability value for the corresponding true class $l(x)$ of pixel $x$, and $x \in \Omega$ such that $\Omega$ indicating the domain of all image pixels. $w(x)$ is the weight for pixel $x$. In paper [9], the weights are pre-calculated by assigning higher values to challenging boundary pixels based on a distance map. The weights are pre-determined and application dependent.

  In our proposed method, we firstly improve the U-net by adding batch normalization (BN) [17] and residual block (RB) [18] to the network layers, as illustrated in Fig. 1(b). BN and RB are well-known techniques to achieve faster convergence and train deeper networks [19]. More importantly, we assign automatically calculated weight to $w(x)$ in equation (1). The weights are pixel-wise values which are iteratively updated in each training iteration for each training image. Calculation of the weight is introduced in section 2.2.

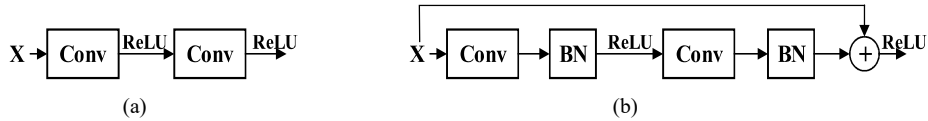

**Fig. 1. (a)** Layer of the original U-net. **(b)** Layer by adding batch normalization (BN) and residual block (RB).

## 2.2   Weighted Cross-entropy Cost Function

In this section, we describe the method for automatically calculating the weight *w(x)* in equation (1).

Object class with a larger number of pixels contributes more to the cross-entropy loss calculation and has larger influence on the gradient values for parameter optimization. Abraham and Khan [20] and Wang et al. [21] have applied dice coefficient loss to address the class imbalance issue. Weight calculated based on the number of pixels per class has also been proposed [6]. Different from these methods that use fixed weight calculations, we propose to calculate the weights dynamically according to the predication performance in each iteration. Our motivation is to increase the contribution from pixels that have larger prediction errors to the loss function calculation. This not only enables the balance of different classes implicitly, but also allows difficult local image regions to be emphasized for model training.

A pixel-wise weight map is generated based on the pixel-wise probability values that are produced in each training iteration. The pixel locations with lower prediction accuracies are assigned to higher weights and vice versa. Hence, the network is able to focus on learning from poorly predicted image regions. The feedback weight is a continuous function that maps the input values to the range of [0.01 1], which is expressed as:

$$w(x) = e^{-log100 \times p_{l(x)}{}^\beta} \qquad (2)$$

In equation (2), larger values of $p_{l(x)}$ indicate higher predicated probability values of the true class, which are assigned to lower weights for calculating the loss function for network backpropagation. Fig. 2 shows the behaviors of the weighting functions by varying the hyperparameter *β* in equation (2). *β* is experimentally determined in section 3. Note that *log100* is used to constrain the minimum weight to be 0.01 instead of 0, which prevents the pixels with high prediction accuracies being completely neglected from training.

Note that the same mapping function in equation (2) is applied to all training images, and mini-batch method [22] is used for parameter optimization. In each batch, a training image with larger poorly predicted regions contributes more than an image with a higher prediction accuracy. This effectively not only balances the image regions but also balance the 'easy' and 'difficult' training examples. This is particularly beneficial for model training based on a small number of training examples with certain bias. We demonstrate this advantage by varying the size of the training data in the evaluation section.

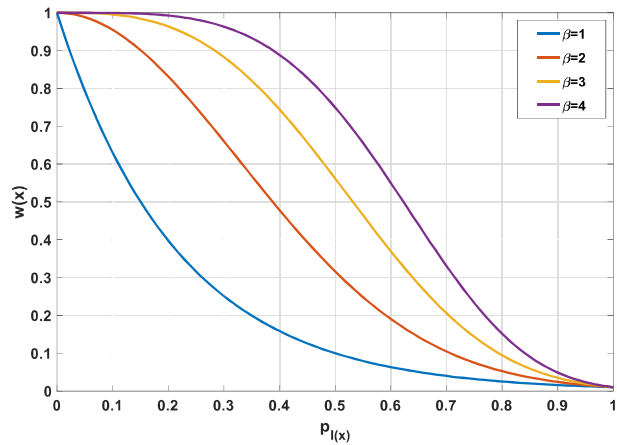

**Fig. 2.** Plot of the mapping function of equation (2) with different values of β.

## 3      Experiments and Results

In this section, we evaluate the proposed method based on T1-weighted brain MRI for segmentation of Midbrain (MB) and Substantia Nigra (SN). Certain quantitative measurements (e.g. volume) of SN has been found to associate with Parkinson disease [23]. However, it is extremely time consuming to annotate it manually and it is challenging to train a machine learning model for automatic segmentation due to the small size of SN. The T1-weighted brain MRI data were acquired in Nottingham University Hospital and was approved by the local ethics committee for this research. The dataset contains a total of 102 subjects with 30 axial image slices each. Experienced radiologist manually selected 3 or 4 slices that contain both the MB and SN, and annotated the contours of MB and SN. This resulted in a total of 310 2D slices for the segmentation evaluation in this paper.

Original U-net, U-net with batch normalization and residual block (BRU-net), BRU-net with feedback weight (FU-net) were compared with each other. The dice coefficient (DC) was used as the evaluation criterion. Note that the separate effects of adding batch normalization and residual block were not tested, as they were normally used simultaneously to achieve better performance. We performed three experiments for each method: randomly selected 200/ 100/ 50 images for training, 10 images for validation and the remaining 100/ 200/ 250 images for testing.

The parameters for model training are listed as follows. The batch size was 5. The optimizer was Adam [24] with learning rate of 0.001. The number of feature channels in the first layer was 16 and doubled in each of the down-sampled layers. The dropout rate was 0.25, and the number of epochs was 400. The number of iterations within each epoch for the three experiments were 40, 20 and 10 respectively (corresponding to experiments with 200, 100 and 50 training images). We evaluated the performances by varying the hyperparameter $\beta$ (equation (2)) from 1 to 4 for the 100 training/200 testing experiment. When $\beta=3$, it achieved the best performance. Hence, $\beta=3$ was used for all the remaining experiments. The main aim of the evaluation is to compare the performances of the proposed improvements rather than achieve an ultimate performance for a particular medical application. Hence, data augmentation was not used.

Table 1 lists the numerical results of the mean DC ± standard deviation (Std) of the three methods by varying the number of training samples. We also report the *P* values of paired t-test by comparing U-net with BRU-net and BRU-net with FU-net respectively.

**Table 1.** Comparison of different methods using different number of training samples. The mean dice coefficient (DC) ± standard deviation (Std) and *P* values of paired t-tests are reported. Numbers in bold indicate the best method that statistically ($P<0.01$) better than other methods.

| Number of training/testing examples | Method | Mean of DC ± Std | |
|---|---|---|---|
| | | MB | SN |
| 200/100 | U-net | 0.9000±0.03 | 0.7095±0.17 |
| | BRU-net | 0.8775±0.14 | 0.7164±0.18 |
| | FU-net | 0.8929±0.05 | **0.7563**±0.15 |
| 100/200 | U-net | 0.8584±0.18 | 0.7022±0.18 |
| | BRU-net | 0.8550±0.16 | 0.7005±0.15 |
| | FU-net | 0.8710±0.15 | **0.7575**±0.16 |
| 50/250 | U-net | 0.8135±0.19 | 0.4831±0.26 |
| | BRU-net | 0.8088±0.15 | 0.6387±0.17 |
| | FU-net | 0.8182±0.20 | **0.7087**±0.24 |
| | | P values of paired t-test | |
| 200/100 | U-net/ BRU-net | 0.0589 | 0.6086 |
| | BRU-net/ FU-net | 0.1260 | 0.0026 |
| 100/200 | U-net/ BRU-net | 0.6050 | 0.8706 |
| | BRU-net/ FU-net | 0.6062 | <0.0001 |
| 50/250 | U-net/ BRU-net | 0.5725 | <0.0001 |
| | BRU-net/ FU-net | 0.3138 | <0.0001 |

It is seen from the results in table 1 that all three methods achieved similar segmentation performance (no statistical significance) for the MB segmentation regardless of the number of training samples. However, for the SN class where the number of pixels is much smaller than the MB class and more difficult to be segmented, the proposed FU-net consistently outperformed the U-net and BRU-net methods for all the experiments with statistical significance. When training using only 50 images, the performance of FU-net remained high (DC=0.7087) which is much higher than the BRU-net (DC=0.6387) and U-net (DC=0.4831).

We also provide some visual examples to demonstrate the advantages of our proposed method. In Fig. 3, we present the segmentation results of an example image based on 50, 100 and 200 training images. Fig. 3(a) and (b) are the original image and ground truth annotation respectively. In Fig. 3(b), the darker region is the MB and lighter region is the SN. Fig. 3 (c), (d) and (e) are the segmentation results for U-net, BRU-net and FU-net respectively using 50 training examples. Some false positives and false negatives can be easily identified for the U-net and BRU-net methods, as indicated by red arrows. Fig. 3(f), (g) and (h) and Fig. 3 (i), (j) and (k) are the results for the three methods using 100 and 200 training examples respectively. Similarly, the FU-net results visually provide more similar outputs to the ground truth image than the other two methods. This is consistent with the numerical results reported in table 1.

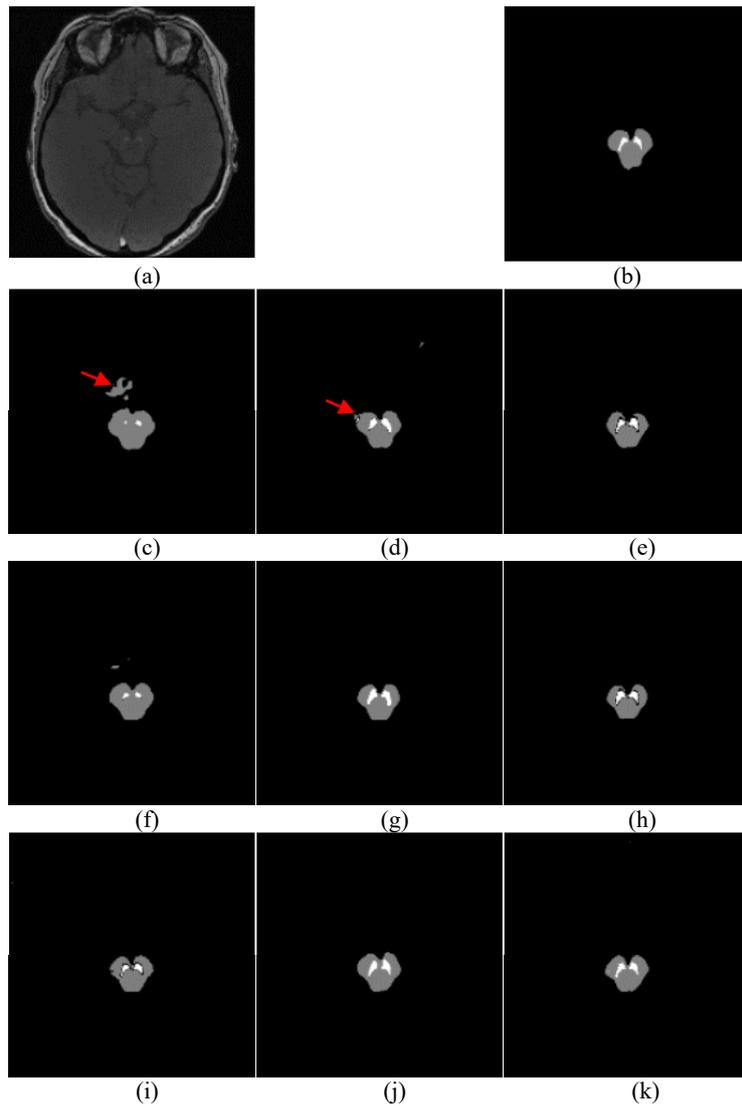

**Fig. 3.** (**a**) The original image. (**b**) The ground truth. (**c**) (**d**) (**e**) Segmentation results by U-net /BRU-net /FU-net respectively using 50 training samples. (**f**) (**g**) (**h**) Segmentation results by U-net /BRU-net /FU-net respectively using 100 training samples. (**i**) (**j**) (**k**) Segmentation results by U-net /BRU-net /FU-net respectively using 200 training samples.

## 4 Conclusion

In this paper, the basic structure of U-net is adopted. We have improved the cost function of U-net by proposing a method to generate dynamic weight. This method enables the prediction accuracy at each training iteration to be used for regionally focused training. The proposed method has been evaluated on a challenging multi-class brain tissue segmentation task. Based on the results, FU-net significantly outperforms the original U-net and an improved version of U-net (BRU-net). We have shown that FU-net is a generic and useful technique for model training with unbalanced class labels and with smaller number of training examples. It can be easily applied to any DCNN based segmentation framework as long as cross entropy is used as the loss function. Future work will focus on method evaluation of different 2D/3D datasets and improvement of the method for tasks with a small number of training samples.

## Acknowledgement

The authors acknowledge Nvidia for donating a graphic card for this research.